\documentclass[conference,usenatbib]{basi}
\usepackage[T1]{fontenc}
\usepackage[british]{babel}
\usepackage[varg]{txfonts}
\usepackage{rotating}
\usepackage{dcolumn}
\begin{document}
\title[Disk geometry of Compact Objects]{Geometry variation of accretion disks of Compact Objects}
\author[P.~S.~Pal et~al.]%
       {P.~S.~Pal$^1$\thanks{email: \texttt{partha.sarathi@bose.res.in}},
       S.~K.~Chakrabarti$^1$ \\
       $^1$S. N. Bose National Centre For Basic Sciences, \\ 
          Sector-III, JD Block, Salt Lake, Kolkata - 700098, India.\\}
%       $^2$S. N. Bose National Centre For Basic Sciences, \\ 
%          Sector-III, JD Block, Salt Lake, Kolkata - 700098, India.\\}
%       $^2$The Second Institute, Address, City, Country}

\pubyear{2013}
\volume{**}
\pagerange{**--**}
%\status{submitted}

\date{Received --- ; accepted ---}

\maketitle
%------------------------------------------------------------------------------%
% abstract and keywords                                                        %
%------------------------------------------------------------------------------%
\label{firstpage}

\begin{abstract}
The Temporal and Spectral variations of black hole candidates during outbursts have been 
reported in several publications. It is well known that during an outburst, the source becomes 
soft in the first few days, and then returns to the hard state after a few weeks or months. In 
the present paper, we show the variation of Comptonization Efficiency (CE), obtained 
from the ratio of the black body photon number to the power-law photon number, 
as a function of time in several outbursts. Since the power-law photons are generated through 
inverse-Comptonization of the intercepted soft photons, the CE is a measure of the 
geometry of the Compton cloud. Our investigation indicated that all the outbursts starts 
with a large CE and becomes very small
after a few days, when the Compton cloud becomes very small to intercept any significant 
number of soft photons. CE returns back to a larger value at the end of the outburst. 
\end{abstract}


%------------------------------------------------------------------------------%
% main text of the paper, using \section, \subsection, \subsubsection          %
%------------------------------------------------------------------------------%
\section{Introduction}\label{s:intro}

\begin{figure}
\begin{center}
\begin{tabular}{p{5.5cm}cp{5cm}}
\raisebox{-\height}
{\includegraphics[viewport=0 150 1000 810,clip=,width=7.0cm]{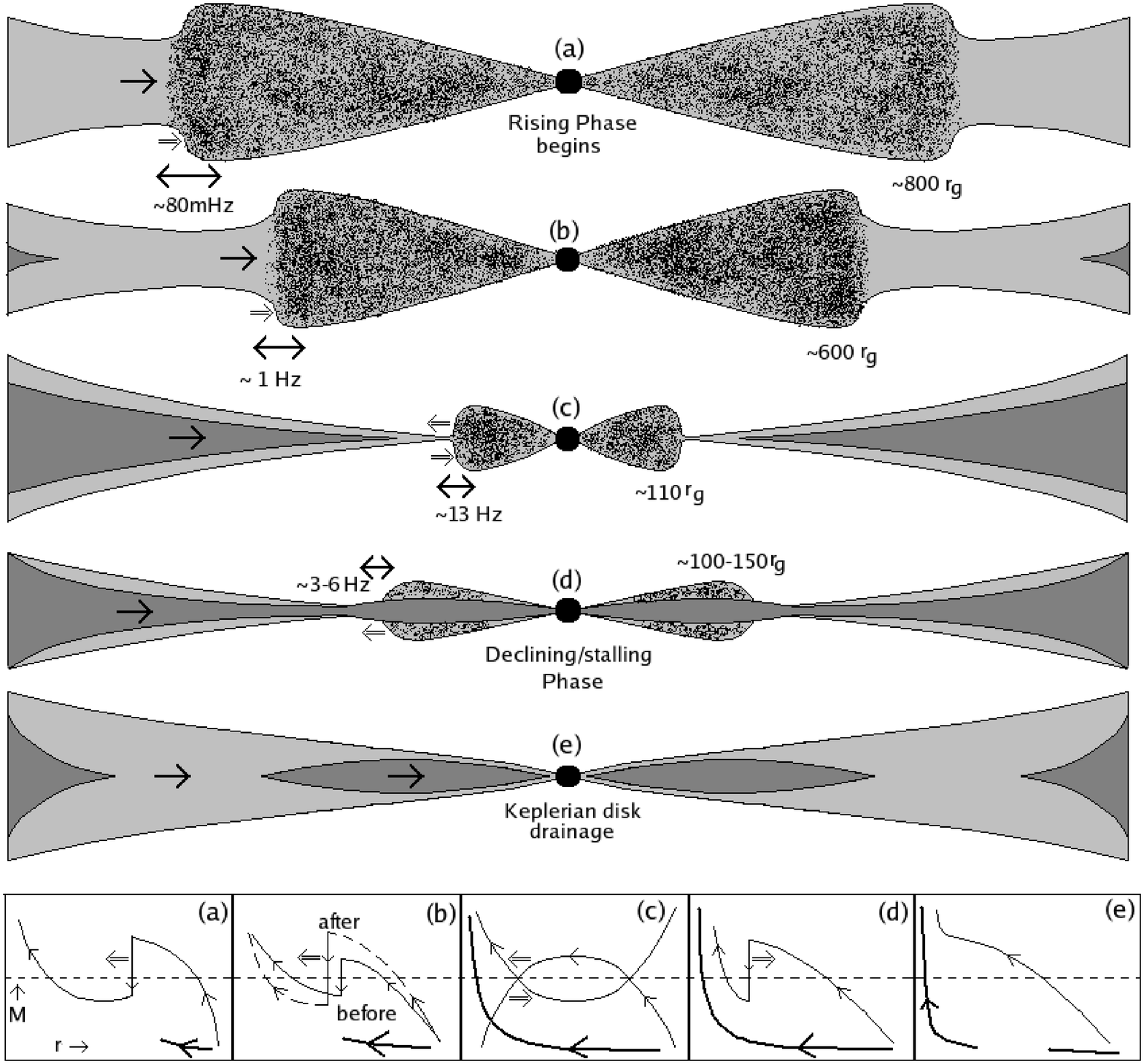}}
& \quad &
\caption{Cartoon diagram of CENBOL geometry variation during outburst of XTE J1550-564. \label{outgeo}}
\end{tabular}
\vskip -1.0cm
\end{center}
\vskip -1.6cm
\end{figure}

It is well known that the  black hole accretion flows in a compact binary system 
typically consist of a Keplerian disk 
which emits soft or low energy photons and a hot Compton cloud which inverse Comptonizes the soft
photons into high energy photons. When the black hole spectral state
changes from the hard state to the soft state and vice versa \citep{C95} 
the Compton cloud must change its shape and temperature: In the soft state when the 
accretion rate in the Keplerian disk is high, the Compton cloud is smaller
and cooler, while in the hard state, when mass accretion rate via Keplerian disk is very low, 
the Compton cloud is larger and hotter. There are various models of the 
Compton cloud in the literature, ranging from the hot Corona, to post-shock region of 
an accreting low angular momentum flow. 
Outburst sources are ideal candidates to study the changes in the size of the Compton cloud. This is
because, the object is known to be in the hard state at the beginning of the outburst, but in a matter
of few days to a few weeks, it changes its states to other states, thereby
giving us a unique opportunity to study the size of the Compton cloud very well. 
Fig.~\ref{outgeo} shows the cartoon diagram of variation of CENBOL geometry during outburst of
XTE J1550-564 \citep{C09}.
From the spectral analysis, 
it appears that the following sequence is typically followed by most, if not all, the outburst sources: 
hard $\rightarrow$ hard-intermediate $\rightarrow$ soft-intermediate $\rightarrow$ soft $\rightarrow$ 
soft-intermediate $\rightarrow$ hard-intermediate $\rightarrow$ hard) \citep{C08, C09}.
Recently, \citet{P11,P12a} showed that the Compton cloud in the variable source GRS 1915+105 changes its size 
as it transits from one variability class to another. We computed a quantity called the 
Comptonizing Efficiency (CE) 
which is the ratio of the power-law photons to the blackbody photons in the spectrum at a given instant. 
We show that CE is very small in softer classes and larger for the harder classes. Within some of the classes, 
there are evidences of rise and fall of the count rates and the spectral slopes in a matter of a few seconds. 
Since the number of hard photons depends on the the optical depth of the Compton cloud, 
we clearly see the change in the optical depth of the Compton cloud in that time scale.

In the present paper, we analyze RXTE data to study the variation of the size of the Compton cloud 
during the outbursts of several black hole candidates. We clearly show that the outburst starts 
with a large Comptonization efficiency,
i.e., a large sized Compton cloud with a poor soft photon source. 
In the rising phase, the cloud becomes smaller and smaller on a daily basis till 
it became minimum when the object went to a soft state. In the declining phase of the outburst, the trend is reversed. 
The black hole candidates we analyzed are GRO J1655-40 and XTE J1550-564.

\section{The Efficiency of Comptonization}

In order to obtain the correct number of soft photons, we first correct for the energy 
dependent absorption by the hydrogen column. We compute the photon numbers emitted in the power-law ($N_{PL}$) 
and the multi-color blackbody components ($N_{BB}$). The ratio $N_{PL}$/$N_{BB}$ will be the
Comptonization Efficiency (CE). This is similar to a `Hardness ratio' but the ranges of the 
hard and the soft photons are automatically selected by the fitting process. 
%This way the influence of the mass of the black hole on energy spectrum is eliminated. 
The number of black body photons are obtained following \citet{maki86} 
within the energy range $0.1$ keV to the best fitted energy range obtained from spectral fitting.
The best fitted energy range is obtained by fitting the spectrum with diskbb model and the 
higher range of the spectrum is obtained from the consideration that the reduced $\chi^2$ value from the resultant
fit must be $\sim 1.0$. This higher energy range of diskbb component varies for different data sets.
Comptonized photons $N_{PL}$ are calculated by using the power-law equation given by 
$P(E)=N(E)^{-\alpha} ,$
where, $\alpha$ is the power-law index and $N$ is the total photons/s/keV at $1$ keV.
%It is reported in \citet{tit94}, that the Comptonization spectrum will have a peak at around $3 \times T_{in}$. 
The power-law is integrated from $3 \times T_{in}$ to $40$ keV to calculate the number of 
Comptonized photons (photons/s). The ratio of the calculated power-law photons and diskbb photons 
will give the parameter CE. This CE represents the geometrical size of CENBOL/ hot electron cloud \citep{P11,P12a}. 
%In softer states the Keplerian rate increases than sub-Kelperian rate. As a result the hot electron cloud 
%is cooled down by Keplerian flow and the size of CENBOL decreases. The number of soft photons 
%increases in softer states and the value of CE decreases. In case of harder states 

\section{Results}

\begin{figure}
\centerline{\includegraphics[angle=270,width=5.75cm]{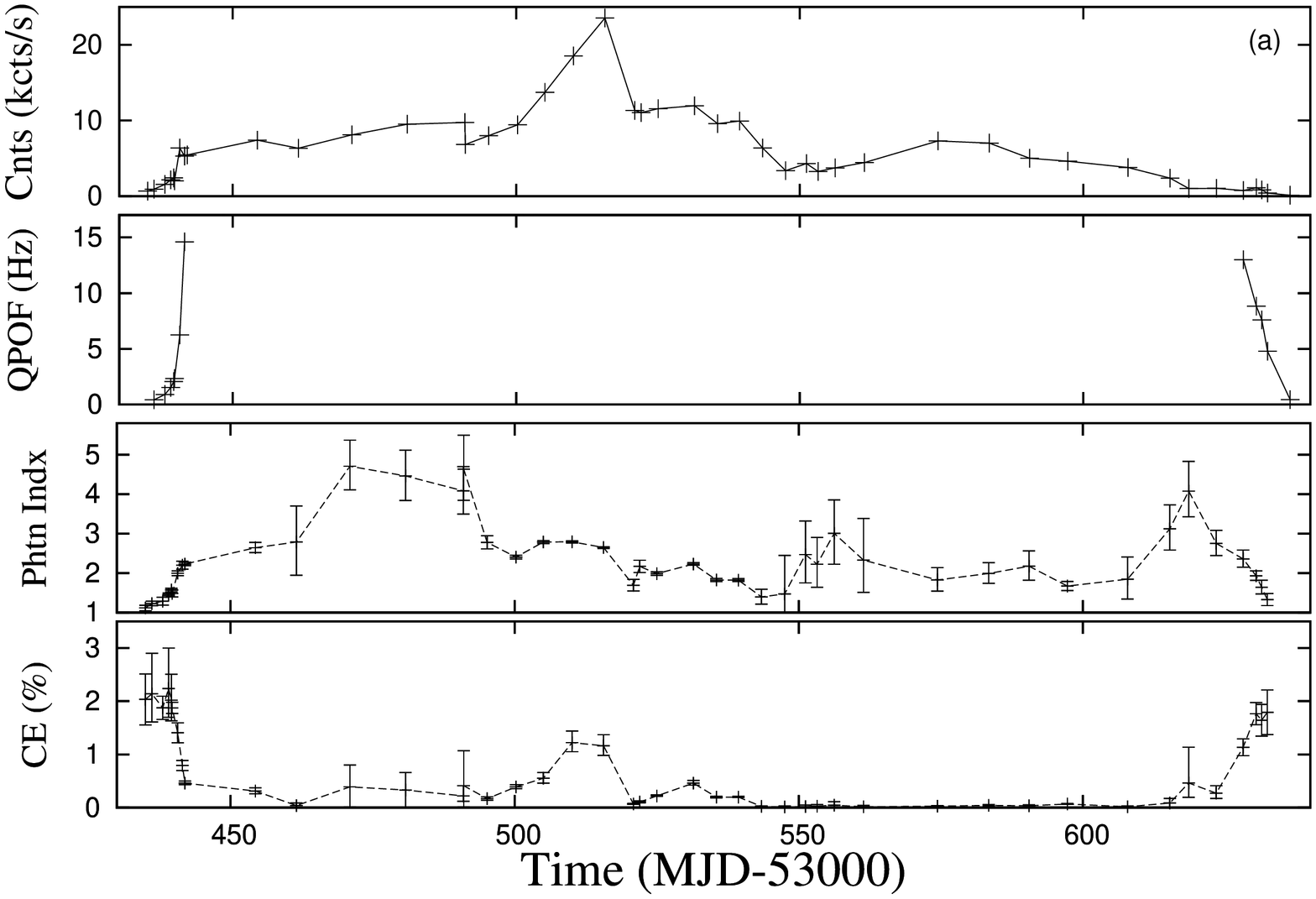} \qquad
            \includegraphics[angle=270,width=5.75cm]{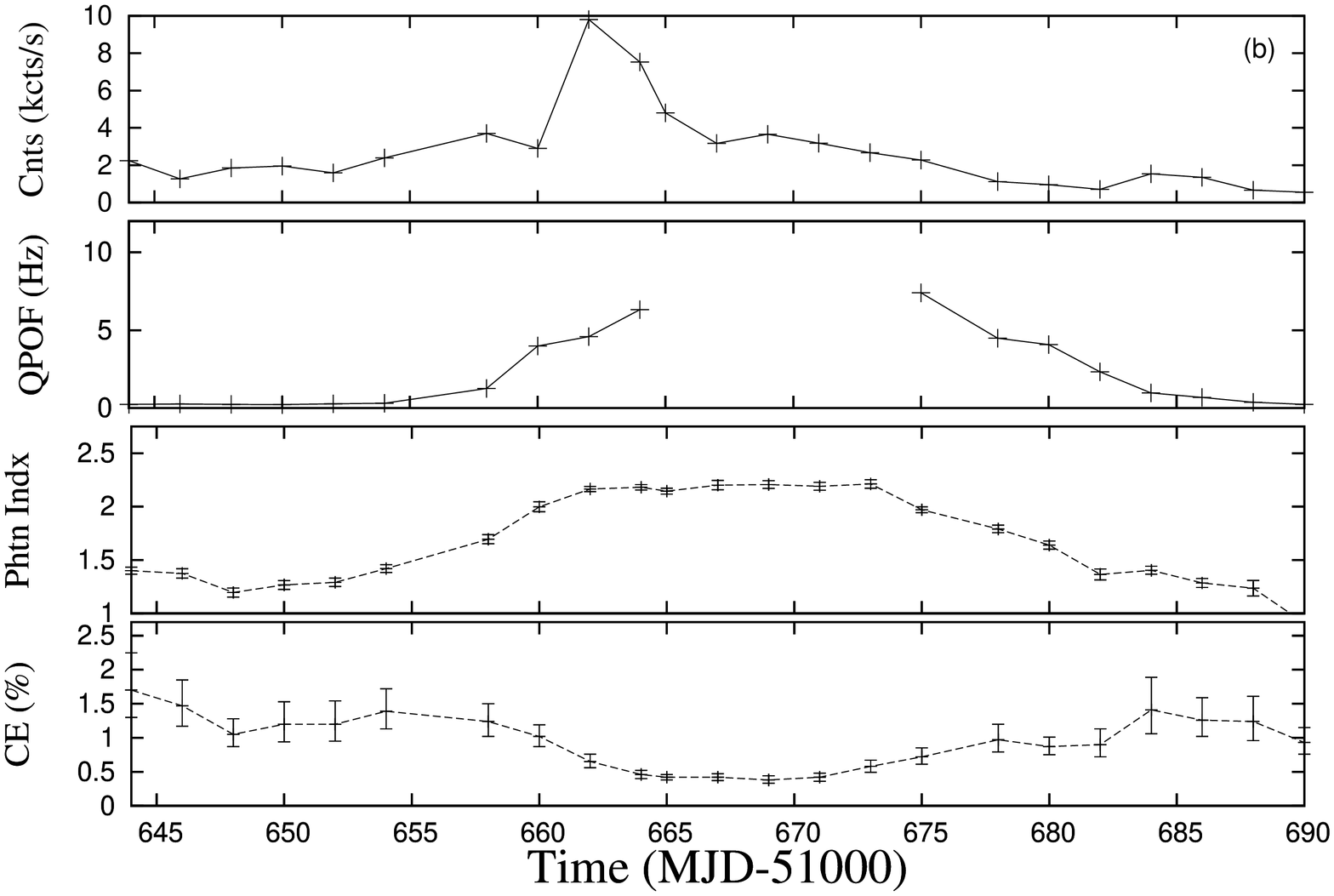}}
\caption{Simultaneous variation of $2.0 - 40.0$ keV counts in kcts/s (upper panel), 
QPO frequency in Hz (second panel), spectral slope (third panel) and the CE (lower panel) of 
(a) GRO J1655-40 during its 2005 outburst and (b) XTE J1550-564 during its 2000 outburst.\label{rslt}}
\end{figure}

\begin{enumerate}
\item {\bf GRO J1655-40}:
The RXTE data during 2005 outburst of GRO J1655-40 is analyzed here and shown in Fig.~\ref{rslt}(a). 
The compact object is analyzed from MJD 53435 (06/03/20 05) to MJD 53632 (19/09/2005). 
During this outburst CE was varying between 2\% and 0.1\%. 
The QPO variation is already reported in \citet{C08}.

\item{\bf XTE J1550-564}: 
The RXTE data during 2000 outburst of the XTE J1550-564 is analyzed from MJD 51644 (10/04/2000) 
to MJD 51690 (26/05/2000). The result is shown in Fig.~\ref{rslt}(b). 
%The light curve looks qualitatively the same as that of the 1998 outburst.
During this outburst, the CE is varying initially between $1.0-1.5$\% but after MJD 51660, the CE 
is reduced to less than $0.5$\%. This indicates that the oscillating shock was still present. 
After a few days, the CE again increased to $\sim 1.5$\% before settling to a $\sim 1$\%. As the shock 
recedes from the black hole, the optical depth initially rises, but then goes down as the CENBOL density
drops. This may be the cause for $CE$ to rise first and then to come down at $\sim 1$\% in the outburst.

\end{enumerate}

\section{Discussions \& Conclusions}
In this paper, we can come to a conclusion that generally speaking, 
all the outbursts start and end with a high Compton cloud size, though
not necessarily the highest optical depth. As the outburst progresses, 
the CE becomes smaller at the peak of the outburst CE has the minimum
value, i.e., the size of the CENBOL is very small. Since CE is a concept
where the hardness ratio is determined from ratio of the dynamically obtained 
black body and power-law photons, it is insensitive to the mass 
of the black hole. This is unlike the usual hardness ratio,
where photons in each energy band could be a mixture of the black body and 
power-law photons. We find that CE in outbursts sources may vary 
from almost $\sim 0.0$ to $\sim 3$. In contrast, the variable source GRS 1915+105 
shows CE between $0.005$ to $0.8$ and has relatively high luminosity (even after 
factoring out the effect of the mass of the black hole), suggesting that it is 
in the soft-intermediate and hard-intermediate states just after the peak of a possible outburst \citep{P12a}.
If so, in future, this source may slow down its activities when the viscosity in the system
is reduced. From \citet{P12b}, we see that the
range of CE is different for different objects. The duration of the outbursts,
the QPO frequency range etc. are also found to be different. There is also a considerable scatter
in CE which may be triggered by outflows from the CENBOL. All these require
a more thorough analysis for a unifying understanding of the outbursts. 
This will be addressed in our future publications.
%------------------------------------------------------------------------------%
\begin{center}
\vskip -0.25cm
{\bf \large Acknowledgments}
\end{center}
\vskip -0.25cm
P. S. Pal acknowledges the SNBNCBS-PDRA fellowship.

\end{document}